# Serverless Computing: Architecture, Concepts, and Applications

Mohsen Ghorbian[1], Mostafa Ghobaei-Arani*[1]

[1] Department of Computer Engineering, Qom Branch, Islamic Azad University, Qom, Iran

*Corresponding Author email: mostafa.ghobaei@iau.ac.ir

**Abstract** Recently, serverless computing has gained recognition as a leading cloud computing method. Providing a solution that does not require direct server and infrastructure management, this technology has addressed many traditional model problems by eliminating them. Therefore, operational complexity and costs are reduced, allowing developers to concentrate on writing and deploying software without worrying about server management. This chapter examines the advantages, disadvantages, and applications of serverless computing, implementation environments, and reasons for its use. Additionally, integrating this computing paradigm with other technologies is examined to address the challenges of managing, securing, and implementing large amounts of data. This chapter aims to provide a comprehensive view of the potentials and limitations of serverless computing by comparing its applications in different industries and examining the future trends of this technology. Lastly, this chapter provides a comprehensive conclusion of the applications and challenges of serverless computing.

**Keywords**: *Serverless Computing , Cloud Computing, Function as a Service, Cloud services, Serverless frameworks*

# 1. Introduction

In recent years, serverless computing has emerged as a transformative paradigm in cloud computing. This innovative approach overcomes many challenges associated with traditional server-based models by eliminating the need to manage servers and infrastructure. Thus, operational complexity and efficiency are reduced, allowing developers to concentrate on writing and deploying code without worrying about server management. Serverless computing solves problems related to capacity management and resource allocation by providing automatic scaling and resource optimization [1,2]. This results in lower costs since users only pay for real-time code execution. More straightforward development and deployment tools significantly reduce development time, and coordination and integration with other systems are improve. Using serverless computing, developers can concentrate on innovation and improving software quality while increasing the security of their code and data due to increased infrastructure security provided by cloud service providers [3,4]. Flexibility facilitates software management and updates, improves real-time processing capabilities and rapid response to events, and simplifies meeting changing business needs. Serverless computing maximizes resource utilization and reduces environmental impact through small and modular functions. Additionally, this model provides extensive support from major cloud providers, providing access to technology for businesses of all sizes and allowing for easy scalability [5]. As serverless computing reduces the costs associated with maintaining and updating infrastructure, it improves productivity and enables businesses to grow and develop. Serverless computing is rapidly increasing due to the many challenges associated with traditional computing. One of the main challenges is managing servers and infrastructure, which requires high technical expertise and considerable time. Traditionally, scaling is handled manually, increasing expenses and additional resources. It is also wasteful to pay fees based on fixed capacity, even when resources are not fully utilized [6]. One of the obstacles to bringing products to market is the longer development and deployment times associated with traditional computing. Furthermore, it is the sole responsibility of development teams to ensure infrastructure security, which can lead to increased security complexity. In light of these challenges, serverless computing has emerged as an efficient and optimal method of increasing productivity and reducing costs [7]. This chapter will examine the advantages, disadvantages, and applications of serverless computing, the implementation environment and reasons for using it. As part of this chapter, we explain how serverless computing benefits business processes, productivity, cost reductions, and customer service levels compared to traditional processing approaches. Furthermore, the integration process of this computing paradigm with other technologies that can address the challenges related to the implementation of this technology, such as big data management, data security, and complex implementation, is examined, and the leading challenges in this field are also examined. Lastly, the chapter provides a comprehensive summary of the potential and limitations of serverless computing by comparing its applications in different industries and offering a glimpse into its future applications.

This chapter structure is as follows: As a starting point, the second part thoroughly examines the necessary prerequisites, such as an Overview of Serverless Computing, Serverless Computing Architecture, Serverless Computing Concepts, and Applications of Serverless Computing. The third part of this chapter describes Serverless Development Environments. The fourth part of the chapter compares serverless computing with traditional computing. The fifth section comprehensively examines Integrating Serverless Computing with Other Technologies. The sixth section examines Future Trends in Serverless Computing. As the last step, the seventh section concludes this explanation.

## 2. Overview of Serverless Computing

Serverless computing is one of the newest and most state-of-the-art data processing and management paradigms in cloud computing environments. It allows developers to develop applications more efficiently and faster by eliminating the need to manage servers. This computing paradigm was supported by major cloud computing providers such as Amazon Web Services (AWS), Google Cloud, and Microsoft Azure [8,9]. It lets developers focus on coding and optimization rather than worrying about hardware and software infrastructure. In addition to providing a ready and scalable infrastructure, serverless computing helps developers respond to customer needs more quickly and efficiently by removing many management and operational challenges from their shoulders. In serverless computing, unlike traditional models that require the purchase, installation, and management of servers, computing resources are allocated dynamically and based on demand rather than requiring the purchase, installation, and management of servers. The costs for any programs are only calculated when running, and users are not charged when they are not used [10]. This paradigm using an economic model of "pay-as-you-go" has significantly reduced operating costs and increased organizational productivity. Additionally, serverless computing can adapt quickly to changes in workloads and data volumes due to its ability to scale automatically. A significant feature of serverless computing is its simplicity and speed in developing and deploying applications [11]. Developers can quickly develop and deploy applications to production environments using serverless cloud services without managing complex infrastructure. This feature is especially attractive for startups and small businesses with limited resources. Furthermore, this computing model, in conjunction with modern tools and frameworks, facilitates software development and reduces the time to market [12]. Serverless computing also helps increase applications' reliability and availability by providing advanced features such as automatic scaling management and synchronization with different load distribution types. Hence, using these features, applications may continue to operate continuously and uninterrupted, irrespective of a sudden increase in workload or an infrastructure issue. The importance of these features is especially evident when it comes to critical applications and services that require high levels of availability [13]. Serverless computing, combined with new technologies such as blockchain, Internet of Things (IoT), and artificial intelligence (AI), enables the creation of intelligent and complex applications and systems. This combination allows businesses to make better decisions and provide better customer services using big data and advanced analytics. Overall, serverless computing has revolutionized the way applications are developed and managed and dramatically changed the future landscape of the IT world by providing new features and capabilities [14].

## 2.1 Serverless Computing Architecture

The serverless computing architecture is a new approach to designing and implementing software programs, eliminating the need to manage servers and physical infrastructure. In this model, developers are only required to write and deploy their code, while cloud service providers handle all infrastructure requirements. This approach includes several components: serverless functions, API gateways, managed database services, and file and data management services [15]. The components of this system are designed to scale and adjust to changing workloads automatically. One of the main components of serverless computing architecture is serverless functions. These functions are small pieces of code that execute only when requested. These functions are usually invoked by API gateways or specific events, such as changes to the database or file uploads. Their main advantage is that they are automatically scalable and do not require manual resource management. This type of serverless function is offered by cloud service providers, including AWS Lambda, Azure Functions, and Google Cloud Functions, which offer developers a wide range of possibilities [16]. Table 1 compares three services providing serverless functions (AWS Lambda, Azure Functions, and Google Cloud Functions) based on ten different parameters.

**Table 1**. Comparing the main features of Serverless Functions from AWS Lambda, Azure Functions, and Google Cloud Functions

| Parameter | AWS Lambda | Azure Functions | Google Cloud Functions |
|---|---|---|---|
| **Execution Time Limit** | • 15 minutes | • 5 minutes (can be increased to 60 minutes in Premium Plan) | • 9 minutes (can be increased to 60 minutes in 2nd Generation) |
| **Memory** | • 128 MB to 10 GB | • 128 MB to 14 GB | • 128 MB to 16 GB |
| **Supported Programming Languages** | • Node.js, Python, Ruby, Java, Go, .NET, PowerShell | • JavaScript, C#, F#, Java, Python, TypeScript, PowerShell | Node.js, Python, Go, Java, .NET, Ruby, PHP |
| **Pricing** | • Based on the number of requests and execution duration | • Based on the number of requests and execution duration | • Based on the number of requests and execution duration |
| **Versioning and Traffic Management** | • Yes | • Yes | • Yes |
| **Event and Trigger Support** | • Yes (Amazon S3, DynamoDB, Kinesis, SNS, SQS, etc.) | • Yes (Azure Event Grid, Storage, Service Bus, etc.) | • Yes (Google Cloud Storage, Pub/Sub, Firestore, etc.) |
| **Monitoring and Logging** | • AWS CloudWatch | • Azure Monitor and Application Insights | • Google Cloud Monitoring and Logging |
| **Local Development and Debugging** | • Yes (AWS SAM Local) | • Yes (Azure Functions Core Tools) | • Yes (Functions Framework) |
| **Integration with Other Services** | • Yes (Integration with AWS services like RDS, API Gateway, etc.) | • Yes (Integration with Azure services like SQL Database, Cosmos DB, etc.) | • Yes (Integration with Google Cloud services like BigQuery, Cloud SQL, etc.) |

| Security and Identity Management | • AWS IAM, AWS KMS, VPC Integration | • Azure AD, Key Vault, VNET Integration | • IAM, Secret Manager, VPC Service Controls |

The API gateway is another key component of the serverless computing architecture responsible for managing the flow of requests and responses between users and serverless functions. API Gateways of cloud providers usually provide features such as authentication, rate limiting, and traffic management. These features make communication between customers and services optimal and secure. Another aspect of serverless computing is managed database services, which allow developers to store and manage their data without managing and maintaining database servers. These services include Amazon DynamoDB, Google Firestore, and Azure Cosmos DB [17]. These databases can handle a large volume of data with high performance, which is particularly important for applications requiring rapid and reliable responses. They are also automatically scalable and can handle large amounts of data. The serverless computing architecture also includes file and data management services, including object storage services such as Amazon S3, Google Cloud Storage, and Azure Blob Storage. These services can store and access large amounts of data in a highly secure and reliable manner. Additionally, these services are commonly integrated into serverless functions and other cloud services to optimize and automate data management processes. Serverless computing architecture provides a dynamic and scalable infrastructure for developing and running software applications by combining these components and services. By eliminating infrastructure and management concerns, developers can focus on innovation and improving user experience while reducing costs, speeding up development, and improving application performance [18]. Figure 1 illustrates the serverless computing architecture.

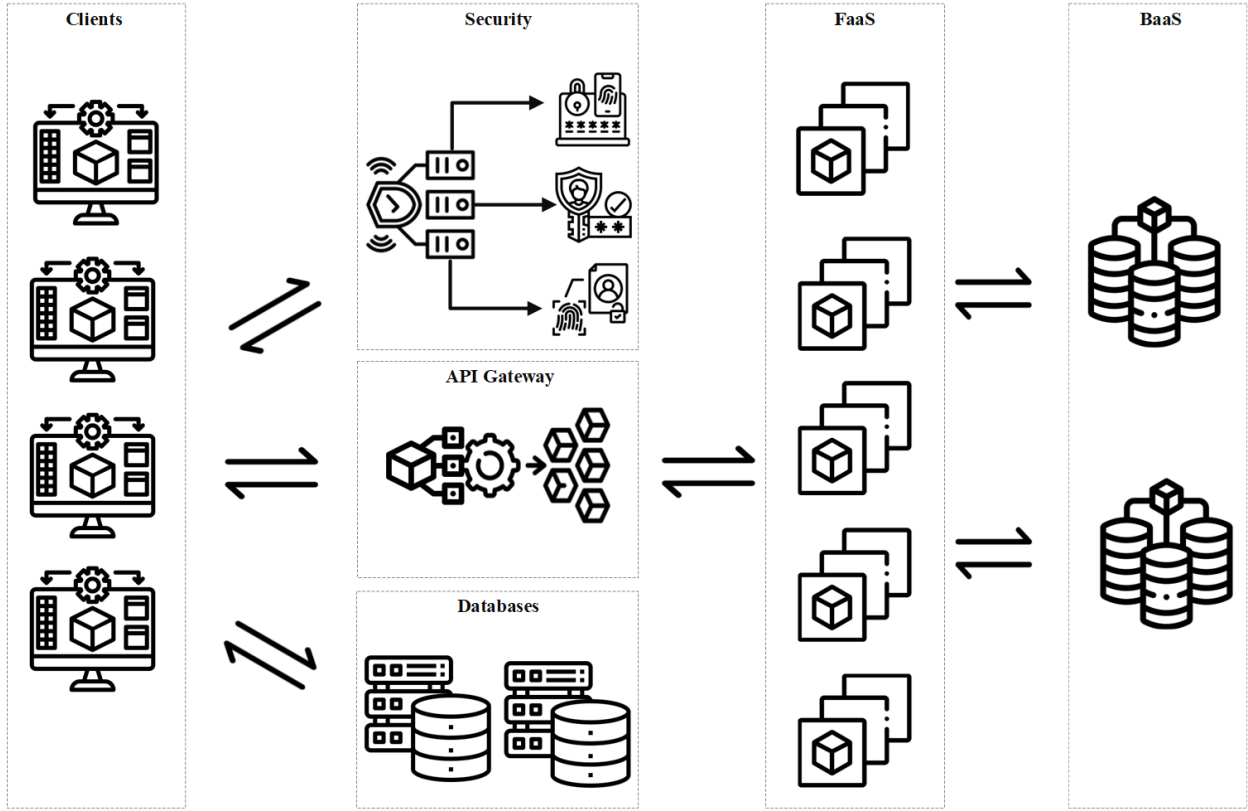

**Figure 1**. the serverless computing architecture.

## 2.2 Serverless Computing Concepts

Serverless computing is one of the most advanced computing technology models, making the program development process faster and more efficient. This model incorporates several key concepts that contribute to optimizing performance and costs. Among these concepts, we can mention "Cold Start," "Function Placement," "Pricing Models," "Auto Scaling," and "Scheduling." These concepts all play an integral role in improving efficiency and reducing costs [19]. The following will provide a detailed analysis of each of these cases.

- **Cold Start:** The term "cold start" describes situations in which a serverless function is invoked for the first time or after a period of inactivity, and there are no active instances of the function in memory when it is called. In this case, the system must restart the execution environment, which includes loading the necessary code and settings. In cases where fast and real-time performance is required, a cold start may cause response time delays and negatively impact the user experience. A cloud service provider may use warm start or pre-warming methods to minimize the effects of cold start [20].

- **Function Placement**: "Function placement" refers to deciding where to execute serverless functions in the cloud infrastructure based on factors including proximity to the end users, server load, costs, and security requirements. The location of functions can significantly affect the performance and cost of a system. For example, if functions are executed near end users, response times can be reduced. Furthermore, properly distributing load among servers can prevent bottlenecks, and resources can be utilized more efficiently. Advanced techniques, such as machine learning, can be used to predict load patterns and optimize where functions are executed [21].

- **Pricing**: "Pricing" in serverless computing refers to pricing models usually determined based on resource usage, such as function execution time, memory consumption, and number of requests. By designing serverless pricing models, users pay for only the resources they use, which minimizes costs. Various pricing models may be offered, including fixed fees per function execution, variable fees based on memory usage and execution time, and volume discounts for large users. Understanding pricing models is essential for optimizing costs and managing budgets, as miscalculations can result in unexpected expenditures [22].

- **Auto-scaling:** "auto-scaling" refers to a system's ability to adjust resources automatically based on the current load demand. In serverless environments, function instances are created and removed according to the volume of requests. Auto-scaling ensures optimal system performance and reduces costs by only utilizing resources when required. Hence, various algorithms are used to perform auto-scaling based on criteria such as request rate, resource utilization, and load prediction. Auto scaling can be performed horizontally (increasing or decreasing the number of instances) or vertically (increasing or decreasing the resource capacity of each instance) [23].

- **Scheduling**: "scheduling" refers to allocating resources and scheduling functions in serverless environments. It includes determining priorities, allocating limited resources, and optimizing the utilization of limited resources. By implementing effective scheduling, productivity can be increased and delays reduced. Resource management in complex environments involves advanced algorithms such as real-time scheduling and multi-objective optimization. Hence, functions are scheduled using real-time scheduling according to their priority and deadline to ensure critical functions are performed at the right time. Additionally, optimized resources can reduce costs and boost overall system efficiency [24].

Combining these concepts can result in improved performance, reduced costs, and increased scalability when running serverless applications. Based on several parameters, it can be helpful to better understand the performance and effects of serverless computing concepts such as cold start, function placement, pricing, auto-scaling, and scheduling. Table 2 shows the main parameters for comparing these concepts, including their advantages, disadvantages, objectives, and applications.

**Table 2**. A comparison of the main concepts based on advantages, disadvantages, objectives, and applications.

| Parameter | Cold Start | Function Placement | Pricing | Auto Scaling | Scheduling |
|---|---|---|---|---|---|
| **Advantages** | • Cost savings during idle times<br>• Improved resource management | • Optimized performance based on function execution location<br>• Reduced response latency | • Pay-as-you-go model<br>• Optimized costs | • Automatic resource adjustment based on needs<br>• Increased efficiency and resource utilization | • Optimal resource allocation<br>• Increased efficiency and reduced latency |
| **Disadvantages** | • Delay during initial startup<br>• Negative impact on user experience | • Complexity in deciding function execution location<br>• Requires advanced techniques for optimization | • Complexity in cost calculation<br>• Potential for unexpected costs | • May not suffice during peak loads<br>• Requires advanced algorithms for load prediction | • Requires complex algorithms for resource management<br>• Issues with prioritizing and scheduling functions |
| **Objectives** | • Cost reduction<br>• Improved system efficiency during idle times | • Improved system performance<br>• Reduced response latency | • Cost reduction<br>• Optimized resource usage | • Maintaining system performance<br>• Enhanced resource efficiency | • Optimized resource usage<br>• Reduced latency and increased efficiency |

| Applications | • Applications with intermittent usage<br>• When cost reduction is a priority | • Distributed systems<br>• When reducing response latency is crucial | • All serverless applications<br>• When cost optimization is a priority | • Applications with variable workloads<br>• When automatic resource adjustment is needed | • Complex systems requiring optimal resource management<br>• Applications with critical timing needs |
|---|---|---|---|---|---|

## 2.3. Applications of Serverless Computing

The serverless computing paradigm has played a significant role in the evolution of various industries. This approach allows businesses to focus on development and innovation without being required to manage servers, as it provides capabilities such as automatic scaling, reduced infrastructure costs, and improved productivity. Various industries, including e-commerce, healthcare, media and entertainment, financial services, transportation and logistics, and education, can benefit from serverless computing by optimizing processes, increasing speed and efficiency, and improving customer service [25]. In the following, each of these applications will be examined in detail.

- **Electronic commerce:** In the e-commerce industry, serverless computing plays a vital role in optimizing the online shopping experience. This technology allows businesses to interact with their customers more effectively and on a larger scale. For instance, product recommendation systems based on user behavior and purchase history can provide more accurate and timely recommendations. Additionally, serverless computing can assist in managing transactions and processing payments, increasing speed and reducing business operational costs [26].
- **Healthcare**: Serverless computing provides smart and efficient healthcare services. Using this technology, hospitals and medical centers can analyze medical data in real-time, which can assist in diagnosing diseases more quickly and providing appropriate treatment to patients. Moreover, serverless computing can be a powerful tool for monitoring patients' conditions and providing immediate alerts [27].

- **Media and entertainment**: In the media and entertainment industry, serverless computing assists in creating and managing multimedia content. This technology is helpful for video and music streaming services, which can process and deliver content around the world using this technology. By reducing infrastructure costs and increasing flexibility, these services can provide better services to their users. Moreover, serverless computing can assist organizations in creating and managing interactive applications and online games [28].
- **Financial services**: By utilizing serverless computing in the financial services industry, banks and financial institutions can improve the efficiency and security of their services. Financial data can be analyzed, transactions can be managed, and intelligent banking services can be provided using this technology. This results in faster, more accurate transaction processing, lower costs, and improved customer satisfaction [29].
- **Transportation and logistics:** Serverless computing can improve supply chain management processes and goods transportation in the transportation and logistics industry. By utilizing this technology, shipping companies can collect and analyze data in real-time regarding the location and condition of goods, which improves planning and reduces costs. In addition, serverless computing can be an effective tool for improving fleet management systems and delivering better customer service [30].
- **Education**: Serverless computing can assist in creating and managing online educational systems. By utilizing this technology, educational institutions can offer courses on demand and online, increasing access to education and improving the learning experience for students. Additionally, serverless computing can analyze educational data and evaluate student performance, improving training processes and providing personalized training [31].

Table 3 compares serverless computing applications in different industries (e-commerce, healthcare, media and entertainment, financial services, transportation, logistics, education, and training) based on criteria such as benefits, challenges, and key applications.

Table 3. A compare serverless computing applications in different industries

| Industry | Benefits | Challenges | Key Applications |
| --- | --- | --- | --- |
| **E-commerce** | • Enhanced online shopping experience<br>• High scalability<br>• Reduced operational costs | • Complexity in managing big data<br>• Data security and privacy | • Product recommendation systems<br>• Transaction management<br>• Payment processing |
| **Healthcare** | • Smart healthcare services<br>• Real-time medical data analysis<br>• Patient status monitoring | • Patient privacy protection<br>• Security of sensitive data | • Faster disease diagnosis<br>• Immediate alerts<br>• Personal health apps |
| **Media and Entertainment** | • Reduced infrastructure costs<br>• Increased flexibility<br>• Better user services | • Managing high volumes of content<br>• Need for extensive user support | • Video and music streaming<br>• Online game management<br>• Interactive apps |
| **Financial Services** | • Increased security and efficiency<br>• Faster transaction processing<br>• Reduced costs | • Countering security threats<br>• Customer privacy protection | • Financial data analysis<br>• Transaction management<br>• Smart banking services |
| **Transportation and Logistics** | • Improved supply chain management<br>• Reduced operational costs<br>• High flexibility | • Managing large datasets<br>  • Need for coordination with various systems | • Real-time tracking and analysis of goods<br>• Fleet management<br>• Improved planning |
| **Education** | • Increased access to education<br>• Enhanced learning experience<br>• Reduced costs | • Managing educational data<br>• Data security and privacy | • Online learning systems<br>• Educational data analysis<br>• Student performance evaluation |

## 3. Serverless Development Environments

The serverless computing development and implementation environment provides developers with tools and platforms to develop and deploy their code without managing server infrastructure. These environments include open-source platforms such as Apache OpenWhisk, Kubeless, Fission, Knative, and OpenFaaS, which are run on Kubernetes and Docker and support a variety of programming languages. Furthermore, closed-source platforms such as AWS Lambda, Google Cloud Functions, Azure Functions, IBM Cloud Functions, and Oracle Functions are managed and integrated with other cloud services. Figure 2 shows these environments, including both open and closed-source environments. These environments provide automatic scaling capabilities, event management, and integration with various services, simplifying and speeding up the development and implementation process. However, the installation, configuration, infrastructure management, and costs associated with increasing the volume of requests can also present challenges, including the complexity of installation and configuration [32]. The following will examine open-source development environments in detail.

- **Apache OpenWhisk**: This platform allows developers to execute small functions in response to various events using a distributed architecture. It is a serverless, event-driven computing platform. The OpenWhisk platform is based on Kubernetes and Docker and is supported by various programming languages, including JavaScript, Python, Swift, and Java. As the platform utilizes the combination and coordination of functions, it can build complex chains of operations and perform automatic scaling.
- **Kubeless:** This platform is an open-source serverless framework that runs on Kubernetes and allows developers to run their functions natively using Kubeless. It fully integrates with Kubernetes, enhancing its functionality management and scalability capabilities. In addition to supporting Python, Node.js, Ruby, PHP, and other programming languages, Kubeless also supports handling events and binding them to functions.
- **Fission**: This open-source serverless platform is built on Kubernetes and designed to develop and run small functions. This platform enables users to develop and run code quickly and supports various programming languages, including Python, Node.js, Go, etc. Fission is event-driven and can respond to HTTP requests, schedulers, and Kafka messages. Fission is easy to use and fast to develop and implement.
- **Knative**: This open-source platform for building and running serverless applications on Kubernetes allows developers to run serverless applications with high performance on Kubernetes. In addition to supporting several languages for developing functions, Knative includes several managed services, such as Knative Serving for executing functions and Knative Evening for managing events.
- **OpenFaaS (Open Function as a Service)**: This is an open-source platform for building and running serverless functions on Kubernetes and Docker. It allows developers to easily create and manage their functions using a straightforward user interface. In addition to supporting Python, Node.js, Go, Ruby, and Java, functions can be designed to respond to

HTTP requests and a scheduler, among other events. One of the outstanding features of OpenFaaS is the ability to automate scaling and use Prometheus for monitoring and scaling functions [33].

The following will examine close-source development environments in detail.

- **AWS Lambda**: AWS Lambda is one of the most popular serverless services provided by Amazon Web Services. The service allows developers to run their code without having to manage servers, and the code is executed in response to events such as changes in S3, updates in DynamoDB, and messages sent to SQS. Lambda scales automatically and supports multiple languages such as Python, JavaScript (Node.js), Java, C#, Go, and Ruby. Full integration with other AWS services, such as API Gateway, S3, and CloudWatch, is one of the prominent features of this service.
- **Google Cloud Functions (GCF)**: GCF is a serverless service from Google Cloud Platform that allows developers to run their code as independent functions. Consequently, these functions are automatically scaled and respond to events such as HTTP requests, changes in Cloud Storage, and Pub/Sub messages. In addition to supporting multiple languages, including Node.js, Python, Go, and Java, Cloud Functions is compatible with other Google Cloud services such as Firestore, Pub/Sub, and Stackdriver.
- **Microsoft Azure Functions:** Microsoft Azure Functions is a serverless service that allows developers to write code as small functions and execute those functions automatically in response to events. The service can respond to events such as HTTP requests, Service Bus messages, and changes to Azure Blob Storage and automatically scales. In addition to supporting languages such as C#, JavaScript (Node.js), Python, Java, and PowerShell, Azure Functions integrates with other Azure services, including Cosmos DB, Event Grid, and Application Insights [34].
- **IBM Cloud Functions**: IBM Cloud Functions is a serverless service built based on Apache OpenWhisk. It allows developers to write and execute their code as small functions. IBM Cloud Functions can respond to multiple events and scale automatically. In addition to supporting JavaScript (Node.js), Python, Swift, and Java languages, it integrates with IBM Cloudant, IBM Message Hub, and IBM Watson services.

- **Oracle Functions**: This serverless service from Oracle Cloud is based on the Fn Project. It allows developers to run their functions based on events and integrate with other Oracle Cloud services. With Oracle Functions, users can write functions in JavaScript, Python, Go, Java, and Ruby, which scale automatically. In addition to Oracle Object Storage, Oracle Autonomous Databases, and Oracle Event Hub, this service is well integrated with other Oracle Cloud services [35].

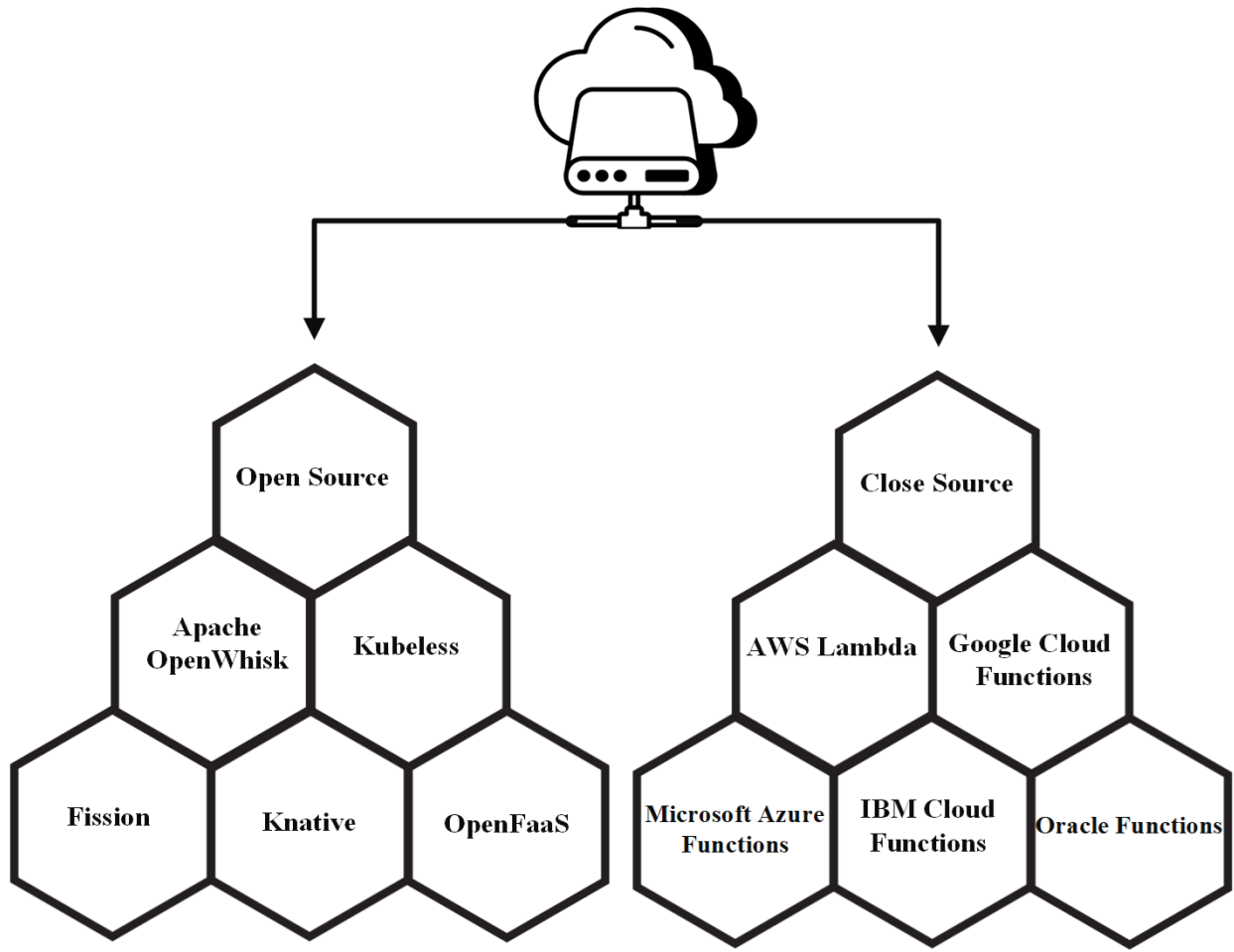

**Figure 2**. Serverless Development Environments: open source and closed source

Depending on the project's requirements and existing infrastructure, these open-source and closed-source platforms can be an appropriate choice based on their features and benefits. The development team can select these tools based on workload, programming language, scalability, and Kubernetes knowledge.

## 4. Serverless Computing vs. Traditional Computing

Comparing serverless computing with traditional computing involves two different approaches. Information technology (IT) teams are traditionally responsible for configuring, maintaining, updating, and scaling the servers and infrastructure necessary to run applications. The process typically requires a high degree of technical expertise and significant human resources and is usually time-consuming and expensive. Alternatively, serverless computing significantly reduces these complications by allowing developers to write and upload their code to cloud environments instead of configuring or managing the supporting infrastructure directly. As a result, organizations can be more productive and drastically reduce infrastructure management costs by automatically configuring, scaling, and managing the required infrastructure as needed [36]. Figure 3 illustrates the differences between serverless computing and traditional computing architectures. The following will examine this difference in detail.

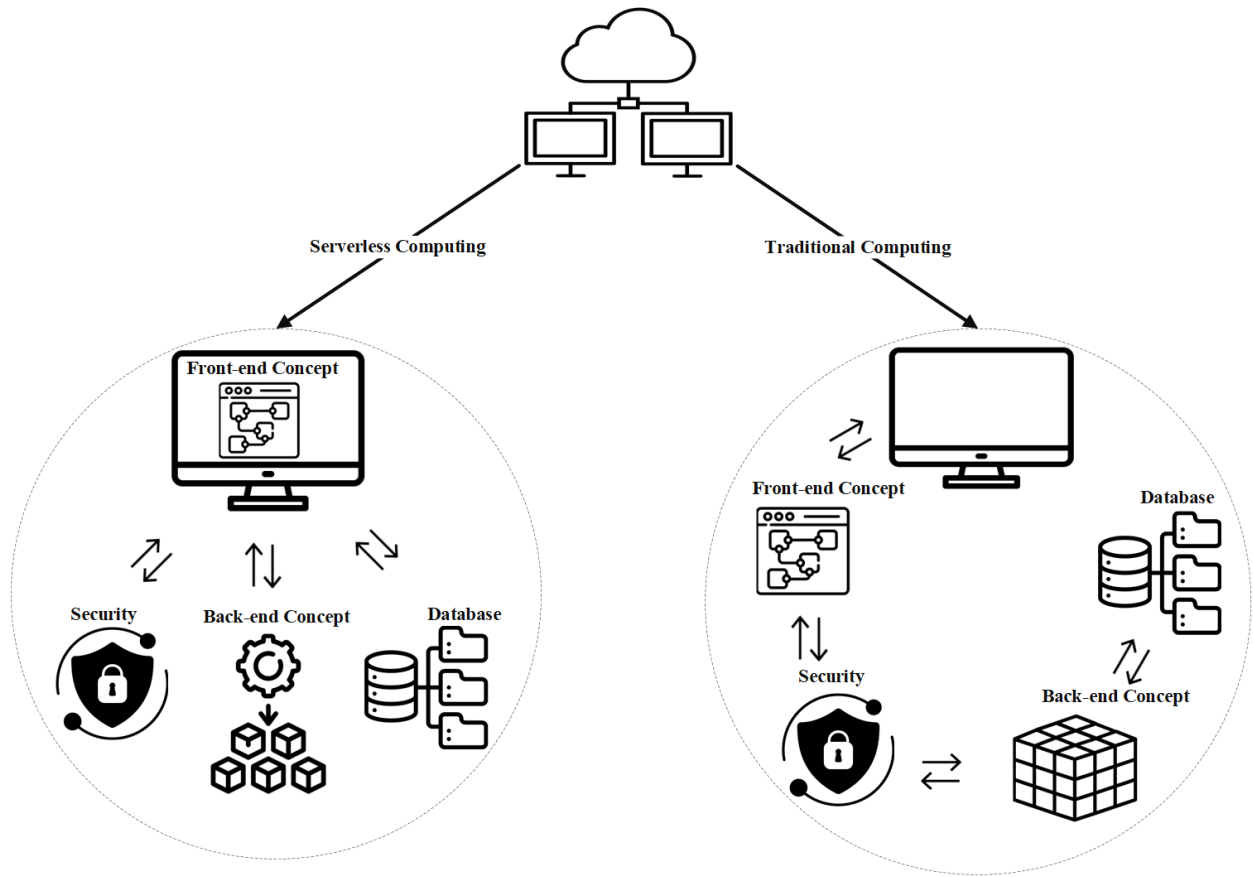

**Figure 3**. Comparison of serverless computing and traditional computing architectures.

- **Infrastructure management:** A traditional computing model entails infrastructure management by information technology (IT) teams, who are responsible for configuring, maintaining, updating, and scaling servers. In contrast, serverless computing transfers these tasks to cloud service providers, which require significant technical expertise and human resources. Developers simply upload their code to the serverless service, and the service manages and scales the infrastructure automatically.
- **Costs:** A traditional computing model typically involves a fixed cost. Organizations must purchase and maintain servers and related infrastructure, regardless of whether fully utilized. These costs include maintenance costs for hardware, electricity, cooling, and manpower. On the other hand, serverless computing uses a pay-as-you-go model, in which fees are incurred only when a particular code is executed. With this model, organizations can significantly reduce costs, especially for programs that run irregularly or in low volumes.
- **Scalability:** The scaling process for a traditional computing system is complex and time-consuming. IT teams must add server resources and make necessary adjustments, which can be time-consuming and costly. With serverless computing, scaling occurs automatically and on demand. Serverless services can be easily and quickly scaled up or down based on changing workloads without requiring manual intervention [37].
- **Development time:** A traditional computing requires much time and energy to set up and configure the infrastructure. Typically, developers are responsible for setting up development and production environments and troubleshooting compatibility issues. In contrast, serverless computing reduces these issues. Developers can focus on writing code and deploy and manage their code using the tools available in cloud services. Doing so reduces the development time, and the product is delivered to the market more quickly.
- **Security:** In the traditional computing, security issues are typically the responsibility of the organization's information technology and security teams. These teams ensure that security updates are made, configurations are secure, and systems are continuously monitored, which requires a considerable amount of time and resources. In serverless computing, service providers are responsible for basic security. They keep their infrastructure secure and apply security updates automatically. Nevertheless, developers are still responsible for their code and data security [38].

- **Flexibility**: Traditional computing requires management of servers and infrastructure, which makes them less flexible. Changes and scaling are typically time-consuming and complex. On the other side, serverless computing provides greater flexibility. In serverless computing, developers can quickly add new functions, implement changes, and respond to business requirements. Serverless services adapt quickly to changes and allow high-scale code execution without manual intervention.
- **Maintenance and support:** In traditional computing, maintenance and support require considerable resources and time. In serverless computing, these tasks have been transferred to service providers responsible for maintaining and updating the infrastructure so developers can focus on improving and developing their applications instead of fixing systems and fixing problems. Security and software updates should also be performed on time. In this model, IT teams can reduce their workload and increase efficiency [39].

Table 4 compares serverless computing and traditional computing based on parameters related to pricing model, server management, scalability, uptime, suitable applications, configuration and setup, security, startup time, programming language support, updates and maintenance, infrastructure control, performance, resource utilization, and response time.

Table 4. A compare serverless computing and traditional computing

| Parameter | Serverless Computing | Traditional Computing |
|---|---|---|
| **Pricing Model** | • Pay-as-you-go | • Fixed and variable costs |
| **Server Management** | • Automatically managed by the service provider | • Manually managed by the IT team |
| **Scalability** | • Automatic scalability | • Manual scalability requiring planning |
| **Uptime** | • Very high (guaranteed by provider SLAs) | • Depends on IT team and infrastructure |
| **Suitable Applications** | • Event-driven and intermittent tasks | • Long-running and complex applications |
| **Configuration and Setup** | • Minimal configuration and setup required | • Extensive configuration and setup needed |

| Security | • Shared responsibility (provider and user) | • Complete responsibility on IT team |
|---|---|---|
| **Startup Time** | • Quick and immediate | • Can be time-consuming |
| **Programming Language Support** | • Multiple popular languages | • Limited to infrastructure capabilities |
| **Updates and Maintenance** | • Automatic and managed by the provider | • Managed by the IT team |
| **Control over Infrastructure** | • Limited | • Complete |
| **Performance** | • May vary depending on traffic | • Predictable and controlled |
| **Resource Utilization** | • Optimized for lower consumption | • Requires precise resource management |
| **Response Time** | • May vary | • Typically, fixed and predictable |

## 5. Integrating Serverless Computing with Other Technologies

Integrating serverless computing with emerging technologies such as blockchains, IoTs, and artificial intelligence (AI) can yield significant performance, scalability, and cost reductions. By integrating serverless computing with these technologies, developers can take advantage of each technology's unique capabilities and develop innovative solutions. The following will provide a detailed analysis of each of these technologies.

- **Integration with blockchain:** Integrating serverless computing with blockchain can facilitate decentralized and secure transactions. Serverless functions can serve as intermediaries between users and blockchain networks, for example, to record transactions, execute smart contracts, and monitor blockchain status. This integration allows for high scalability and low costs, as functions are executed just when needed, reducing permanent servers' fixed costs. Further, serverless computing facilitates processing transactions and delivering results to users more quickly and efficiently [40].
- **Integration with the Internet of Things (IoT)**: The serverless computing approach is naturally compatible with the Internet of Things (IoT), as many IoT devices need to process small amounts of data and react to various events on a timely basis. A serverless function can process incoming data from IoT devices, execute application logic, and store results. In this model, developers can develop IoT applications that are scalable and efficient since computing resources are dynamically allocated according to device needs. In addition, serverless computing can quickly respond to changes in the number of connected devices necessary to handle IoT workload [41].

- **Integration with artificial intelligence (AI)**: Implementing complex algorithms and machine learning models at a large scale is possible through the effective combination of artificial intelligence and machine learning (ML) with serverless computing. Using serverless functions to process data, run artificial intelligence models, and present results to users or other systems is possible. This integration is particularly useful for applications requiring real-time data analysis, such as image recognition, audio analysis, and prediction. The developers can divide their AI models into small, discrete functions and run them in parallel, which increases processing speed and reduces costs [42].

A significant benefit of integrating serverless computing with these technologies is the increased level of security that can be achieved. As part of blockchain, serverless functions can serve as secure intermediaries for smart contract execution, ensuring that transactions are executed securely and transparently. In the Internet of Things, serverless functions can analyze incoming data and identify security threats in real-time. Hence, using artificial intelligence, these functions can implement sophisticated security models to detect intrusions and attacks, thereby assisting in protecting data and systems. Despite the many benefits, integrating serverless computing with these technologies presents challenges. These challenges include concurrency management in blockchains, bandwidth and latency issues in the Internet of Things, and the need for high computing resources in artificial intelligence. To address these challenges, developers must utilize appropriate architectures and best programming practices. Advanced techniques such as sharding in blockchain, data compression in IoT, and optimizing AI models can help improve performance and reduce problems. Additionally, the integration process can be optimized if the right cloud service provider is chosen and the native capabilities of each platform are utilized.

## 6. Future Trends in Serverless Computing

The field of serverless computing is rapidly evolving and expanding, and future predictions and trends predict significant changes and numerous innovations. The adoption and use of serverless computing in various organizations and businesses is one of today's most significant trends. The advantages of flexibility, scalability, and cost reduction have led many companies to utilize this computing model. The use of serverless computing is expected to become an integral part of software development in the near future, and many applications and services will be developed using this technology. Serverless computing is also being improved and developed regarding tools and services. Cloud service providers such as Amazon Web Services, Google Cloud, and Microsoft Azure are constantly upgrading their platforms with new features and capabilities. Among the features included are advanced management and monitoring tools, performance optimization, and enhanced security. In addition, more developers are developing and managing serverless functions using open-source frameworks and libraries, increasing efficiency and reducing implementation complexity. In addition, serverless computing is increasingly integrating with advanced technologies such as artificial intelligence, machine learning, and the Internet of Things. By incorporating these technologies, more innovative and efficient solutions can be developed. For example, serverless functions can run machine learning models and process IoT data, facilitating data analysis and real-time decision-making. The development of more complex and intelligent systems is predicted to occur as a result of these trends in the future. A significant challenge to address in the future is improving management and reducing cold start delays. As serverless computing use increases, the need for faster startup times of functions will become increasingly apparent. Various techniques are considered, including preheating functions, using lighter containers, and optimizing computing infrastructure. These improvements will help reduce delays and improve the user experience.

## 7. Conclusion

Serverless computing has solved many problems in traditional models by providing a solution that does not require direct management of servers and infrastructure. In this way, operational complexity and efficiency are enhanced, allowing developers to focus on writing and deploying code rather than managing servers. This article examines the advantages, disadvantages, and applications of serverless computing, the environments in which it can be implemented, and the reasons for its use. By comparing its applications in several industries and analyzing its future trends, this paper strives to analyze the potential and limitations of serverless computing comprehensively. The article also discusses technological developments in integrating serverless computing with other technologies, as well as recent trends that demonstrate the effectiveness of this technology.